\documentclass{aa}  

\usepackage{graphicx}
\usepackage{txfonts}
\begin{document} 

   %\title{Efficient simulation of Fourier-filtering wavefront sensors with elongated guide stars}
   \title{Expected performance of the Pyramid wavefront sensor with a laser guide star for 40 m class telescopes.}
   %\titlerunning{Test}

   \author{F. Oyarzún
          \inst{1}
          V. Chambouleyron \inst{2}
          B. Neichel\inst{1}
          T. Fusco \inst{3,1}
          \and
          A. Guesalaga \inst{4}
          }

   \institute{Aix Marseille Univ, CNRS, CNES, LAM, Marseille, France\\
              \email{francisco.oyarzun@lam.fr}
         \and
            University of California Santa Cruz, 1156 High St, Santa Cruz, USA\\
         \and
             DOTA, ONERA, Université Paris Saclay, F-91123 Palaiseau, France\\
        \and
            Department of Electrical Engineering, Pontificia Universidad Católica de Chile, 4860 Vicuña Mackenna, 7820436 Santiago, Chile
             }

   \authorrunning{F. Oyarzún}
   \date{\today}

% \abstract{}{}{}{}{} 
% 5 {} token are mandatory
 
  \abstract
  % context heading (optional)
  % {} leave it empty if necessary  
   {The use of artificial Laser Guide Stars (LGS) is planned for the new generation of giant segmented mirror telescopes, to extend the sky coverage of their adaptive optics systems. The LGS, being a 3D object at a finite distance will have a large elongation that will affect its use with the Shack-Hartmann (SH) wavefront sensor.}
  % aims heading (mandatory)
   {In this paper, we compute the expected performance for a Pyramid WaveFront (PWFS) Sensor using a LGS for a 40 m telescope affected by photon noise, and also extend the analysis to a flat 2D object as reference.}
  % methods heading (mandatory)
   {We developed a new way to discretize the LGS, and a new, faster method of propagating the light for any Fourier Filtering wavefront sensors (FFWFS) when using extended objects. We present the use of a sensitivity model to predict the performance of a closed-loop adaptive optic system. We optimized a point source calibrated interaction matrix to accommodate the signal of an extended object, by means of computing optical gains using a convolutional model.}
  % results heading (mandatory)
   {We found that the sensitivity drop, given the size of the extended laser source, is large enough to make the system operate in a low-performance regime given the expected return flux of the LGS. The width of the laser beam, rather than the thickness of the sodium layer was identified as the limiting factor. Even an ideal, flat LGS will have a drop in performance due to the flux of the LGS, and small variations in the return flux will result in large variations in performance.}
  % conclusions heading (optional), leave it empty if necessary 
   {We conclude that knife-edge-like wavefront sensors, such as the PWFS, are not recommended for their use with LGS for a 40 m telescope, as they will operate in a low-performance regime, given the size of the extended object.}

    
   \maketitle
%
%-------------------------------------------------------------------
%\input{Introduction}
%

\section{Introduction}
The next generation of Extremely Large Telescopes (ELTs) will offer unprecedented opportunities for ground-based observations. These telescopes are three to four times larger than their predecessors, providing greater resolving power that enables the detection of finer structures than previously possible. This makes ELTs a promising option for direct imaging of exoplanets and studying their atmospheric composition, potentially leading to the detection of biomarkers \citep{2013ApJ...764..182S}. The light-gathering capability of these new telescopes will be an order of magnitude greater than the previous generation, allowing for the observation of more distant and faint objects, needed to study the early stages of the universe \citep{2007Msngr.127...11G}.

The resolving power of the ELTs will be limited by the atmospheric coherence length $r_0$, which is typically between 10-15 cm. Without atmospheric compensation, ELTs would perform no better than a home telescope. Adaptive optics (AO) is used to overcome this limitation \citep{1977JOSA...67..360H}. AO consists of three main components: a wavefront sensor (WFS), which measures phase aberrations introduced by the atmosphere; a deformable mirror (DM), which corrects these disturbances by deforming its surface; and a real-time computer (RTC), which processes the measurement from the WFS and sends the corresponding signal to the DM at high speed. 

To measure atmosphere distortions, a guide star is necessary. Laser guide stars (LGS) have been used for over 30 years to compensate for the lack of natural guide stars (NGS) bright enough to provide good sky coverage for AO systems \citep{1991Natur.353..141P}. LGSs are generated using a laser to excite sodium atoms present in a layer about 20 km thick at approximately 90 km above sea level \citep{1985A&A...152L..29F}. Due to beam divergence, atmospheric conditions, and the thickness of the sodium layer, the laser beacon in the sky is a cylindrical volume with a width in the order of one arcsec and a height of 20 km, making it a 3D object.

The Shack-Hartmann (SH) wavefront sensor is a popular choice for measuring wavefront aberrations. The SH WFS is a focal plane sensor that measures the gradient of the incoming phase of the wavefront. It uses a grid of micro-lenses, with each lens sampling a portion of the wavefront and producing an image of the source. The position of each image is proportional to the average gradient of the portion of the incoming phase of the wavefront. However, for a 40 m telescope with an 80 x 80 subaperture SH WFS considering a side launch telescope, the LGS spot is four times wider and up to sixty times larger than the diffraction-limited spot of each subaperture. This means that to correctly sample the laser spot, the detector of the SH must have a large number of pixels (e.g. 1600 x 1600) \citep{fusco2019story}. This wavefront sensor is currently being used in the design of some of the ELT first light instruments, such as the High Angular Resolution Monolithic Optical and Near-infrared Integral field spectrograph (HARMONI) \citep{2016SPIE.9908E..1XT} and the Multiconjugate adaptive Optics Relay For ELT Observations (MORFEO) \citep{2022SPIE12185E..14C}.

The pyramid wavefront sensor (PWFS) \citep{1996JMOp...43..289R} is a pupil plane WFS from the family of the Fourier Filtering WFS (FFWFS). Its working principle is similar to the Foucault knife-edge test, but instead of blocking part of the light, the PWFS uses a glass pyramid to split the light in the focal plane and generate four images of the entrance pupil each of which with a specific intensity pattern that encodes phase information. Given the difficulties of the SH WFS with LGS, the PWFS has been proposed as an alternative \citep{2010SPIE.7736E..57L, pinna1a2011pyramid, 2013aoel.confE..15Q, 2015aoel.confE..37B, 2016SPIE.9909E..6BE}, given its higher sensitivity and lower demand on pixels. An equivalent 80 x 80 subaperture PWFS would need a detector no bigger than 240 x 240 pixels, making it possible to use fast, low-noise detectors. 

Modulation of the PWFS is a commonly used technique for NGS, where a known oscillating aberration (typically tip-tilt) is introduced which allows adjusting the properties of the PWFS: an increase in modulation radius gives a higher dynamic range at the cost of lower sensitivity \citep{2004OptCo.233...27V, 2016Optic...3.1440F}. The integration time of the detector has to be an integer multiple of the period of the oscillation. This produces a signal equivalent to having many incoherent point sources arranged in a circle (assuming circular modulation, a static atmosphere during integration and no anisoplanatism).

The SH WFS with LGS for a 40 m telescope will need a detector with too many pixels to cope with the dynamic required to sense LGS spots, affecting the associated sensitivity. We were therefore interested in investigating the performance of the PWFS when using an LGS, given its higher sensitivity when using an NGS and its lower demand on pixels. The performance will be measured as the Strehl ratio obtained in an AO loop for different return fluxes of the guide sources. To do this, we first have to understand the properties of the LGS, and how this artificial star shapes the signal that we measure with the PWFS. As the computing requirements are high (in terms of memory and time) we had to develop new techniques that allowed us to simplify the simulations. These simulations were based on the end-to-end physical optics models from OOMAO \citep{2014SPIE.9148E..6CC}.

The main objective of this work is to study the performance of the PWFS for different sources: NGS, LGS-2D and LGS-3D, and different telescope sizes from 8 to 40 m. The main focus is to test the influence of photon noise in closed-loop operation and to compare the end-to-end (E2E) results with predictions using linear models, and finally be able to compute the expected performance for a 40 m telescope.

One of the main difficulties for computing the performance was the size of the simulations. Considering a side launch LGS, by using a geometrical approach it is possible to compute that the extension of its image in a 40 m telescope is about 20 arcsec, taking into account both the angular size and the depth of field. If using 2.4 pixels per $\lambda/D$ (i.e. 1.2 x Shannon), and an observing wavelength of $589 \, nm$, those 20 arcsec correspond to just under 16,000 pixels. Leaving space for diffraction or atmospheric effects, the matrices that would be needed to propagate have approximately 20,000 x 20,000 complex, double-precision entries for each sample of the LGS. The interaction matrix for an 80x80 deformable mirror would need around 11,000 frames to be computed. Even for good computers, this calibration process might take months or even years.

To achieve the goal of computing the performance of the 40 m telescope, in Sect. \ref{sec:MandS} we present the mathematical formalism we use to process the raw data from the PWFS, and introduce an analytical model that we will use to predict the performance for the 40 m telescope, instead of having to do the full end-to-end simulations. This model requires the interaction matrix of the system, therefore Sect. \ref{sec:LGS} is about how we simulate the LGS, such that we are able to build an end-to-end interaction matrix. Sect. \ref{sec:IMat_cal} presents some of the issues and alternatives for the computation of the interaction matrix both in simulation and for a real telescope. Sect. \ref{sec:CMod} is about how to use a convolutional model to optimize the interaction matrix for the LGS, from the one calibrated using a point source. Finally, in Sect. \ref{sec:closed_loop} we show end-to-end closed-loop simulations of smaller telescopes (8 and 16 meters) to validate the predictions of the analytical noise model. With the validated model, we are then able to extrapolate the results and compute the expected performance of the AO loop for the 40 m telescope.

\section{Data processing and noise propagation}
\label{sec:MandS}

\subsection{Signal and reconstruction}

The framework we used for the signal processing of the PWFS in this work is the one presented in \citet{2023A&A...670A.153C}. Given an input phase $\phi$, we processed the raw signal from the PWFS $I(\phi)$ to obtain the reduced intensities $\Delta I(\phi)$. For the reference intensity $I_0$ we used the signal of the PWFS corresponding to a flat wavefront. We built the interaction matrix $\mathcal{D} = [\delta I(\phi_1),\dots, \delta I(\phi_N))]$ using the push-pull method inputting an orthogonal basis $[\phi_1,\dots,\phi_N]$ in the phase space corresponding to Karhunen-Loève modes. The full notation can be found in App. \ref{app:framework}

The reconstructor can then be obtained as the pseudo-inverse of $\mathcal{D}$ as $\mathcal{D}^\dagger  = (\mathcal{D}^t \mathcal{D})^{-1} \mathcal{D}^t$. Assuming a small phase regime and the linearity of the PWFS, the modal reconstruction of the phase $\phi'$ can be obtained with the following matrix-vector multiplication

\begin{equation}
    \phi' = D^\dagger \, \Delta I(\phi).
    \label{eq:fst_recon}
\end{equation}

\subsection{Noise propagation}

Noise in the AO loop is given by two distinct terms: read-out noise (RON) and photon noise. The residual variance due to noise for each corrected mode is given by the sum of both noise contributions. In this work, we will use the analytical model developed in \citet{2023A&A...670A.153C}, and the specific notation can be observed in App. \ref{app:noise}. As the detectors needed for the PWFS are small, it is possible to use ultra-low noise detectors with sub-electron RON \citep{2011aoel.confE..44G}, meaning that we can neglect read-out-noise, therefore it is possible to assume that the only contribution to noise is photon noise. Even if detector noise would have a significant impact, photon noise is a fundamental limit that is independent of the technology being used. With the analytical model it is possible to compute a sensitivity to photon noise $s_\gamma(\phi_i)$ (See \citet{2023A&A...670A.153C}, Eq. 23 or App. \ref{app:noise}, Eq. \ref{eq:s_gamma}), which encodes the robustness of the system to photon noise when measuring the amplitude of mode $\phi_i$ with the PWFS. If we have a frame with $N_{ph}$ photons, the total residual variance introduced by photon noise at that measurement is

\begin{equation}
    \sigma_\gamma^2 = \sum_{i = 1}^n \frac{1}{N_{ph} \, s_\gamma^2(\phi_i)}.
    \label{eq:sigma_gamma}
\end{equation}

\begin{figure}
    \centering\includegraphics[width = 0.49\textwidth]{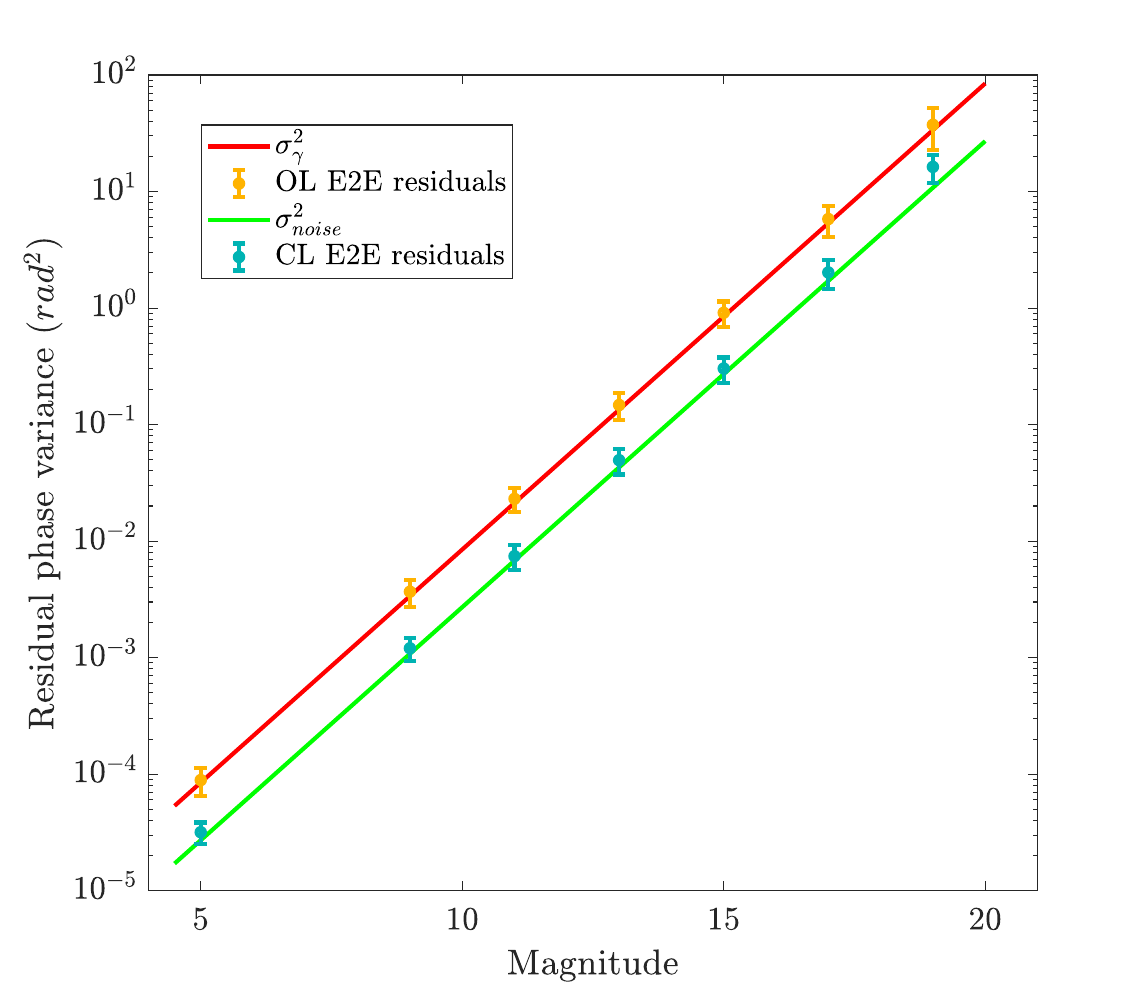}
    \caption{Evolution with respect to the magnitude of the guide star of the residual phase variance due to photon noise in open and closed loop for an 8 m telescope with a $4\lambda / D$ modulated NGS. The solid line corresponds to the residual phase variance $\sigma^2_{noise}$ using sensitivity analysis from equations \ref{eq:sigma_gamma} and \ref{eq:sigma_noise}, and the markers correspond to the mean of 200 end-to-end iterations, with the errorbar the standard deviation of the residual variance. For reference, at magnitude 10, $N_{ph} = 4.5 \times 10^4$ photons.}
    \label{fig:NGS_sens_pred}
\end{figure}

To compare this approximated model with E2E simulations, we simulated 200 realizations with no atmosphere (i.e. a flat wavefront) of an open loop AO system with no controller for a $4 \, \lambda/D$ modulated NGS with magnitudes from 5 to 20 to observe the effects of photon noise in the residual variance. We found good agreement between the E2E simulation and the sensitivity model, as it can be seen in Fig. \ref{fig:NGS_sens_pred}, with the predicted residual variance as the red line and the E2E simulations as the orange markers.

We needed to see if our noise model works in a closed loop, taking into account its temporal properties. The controller used was a discrete integrator in the feedback path with gain $\alpha$. The dynamics of the DM were modeled as a Zero-Order Hold (ZOH) and the WFS as a ZOH with a time delay of one period $T$. An additional time delay of one period was assumed for the computation of the signal. Taking $\alpha = 0.3$ and a sampling frequency $F = 1 \, kHz$, we can integrate the magnitude squared of the noise transfer function over the bandwidth to obtain the total noise $\sigma^2_{noise}$ that is propagated through the AO loop. Solving the integral we get

\begin{equation}
    \sigma^2_{noise} = \delta \sigma_{\gamma}^2,
    \label{eq:sigma_noise}
\end{equation}

with $\delta = 0.33$ (See the details in App. \ref{app:CTheory}). To compare the model with the E2E simulations, we simulated 200 realizations of a closed loop for a flat wavefront. We found good agreement with the sensitivity analysis combined with the control theory, as can be observed in Fig. \ref{fig:NGS_sens_pred}. The light green line represents the residual variance predictions and the E2E results as the bluish green markers, showing good agreement between simulation and theory. The uppermost point of the closed loop E2E simulation deviates from the expected behavior. This might be because of three reasons: the non-linearities of the pyramid decrease the sensitivity, which increases the overall noise propagated, photon noise sensitivity assumes that the illumination pattern in the detector is similar to the reference intensity, therefore for large residual phase variances, the sensitivity might not accurately predict the propagation of noise, and the non-linearities of the pyramid having an effect on the control loop which were not taken into account when computing the NFT. The first two affect both the open and closed loop cases, therefore the latter is the most plausible explanation of the deviation of the uppermost point of the closed loop.

To be able to compute the sensitivity of the system, we have to compute the interaction matrix. The next section is about how we simulated the LGS, such that we can then build an end-to-end interaction matrix.

\section{LGS simulation}
\label{sec:LGS}
\subsection{LGS geometry}

\begin{figure}
    \centering
    \includegraphics[width = 0.4\textwidth]{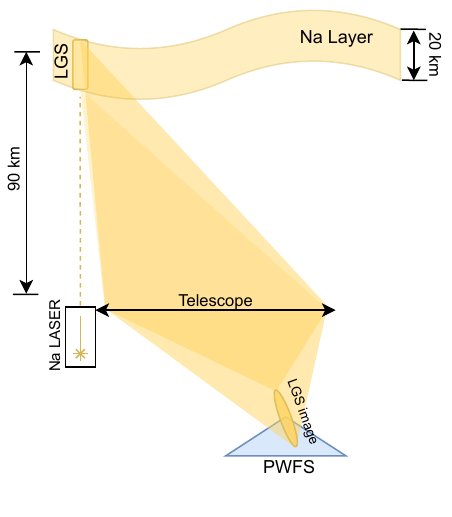}
    \caption{Schematic of the LGS and the image formation. The drawing is not to scale and proportions were altered for ease of understanding.}
    \label{fig:LGS_schematic}
\end{figure}

\begin{figure*}[t]
    \centering
    \includegraphics[width = \textwidth]{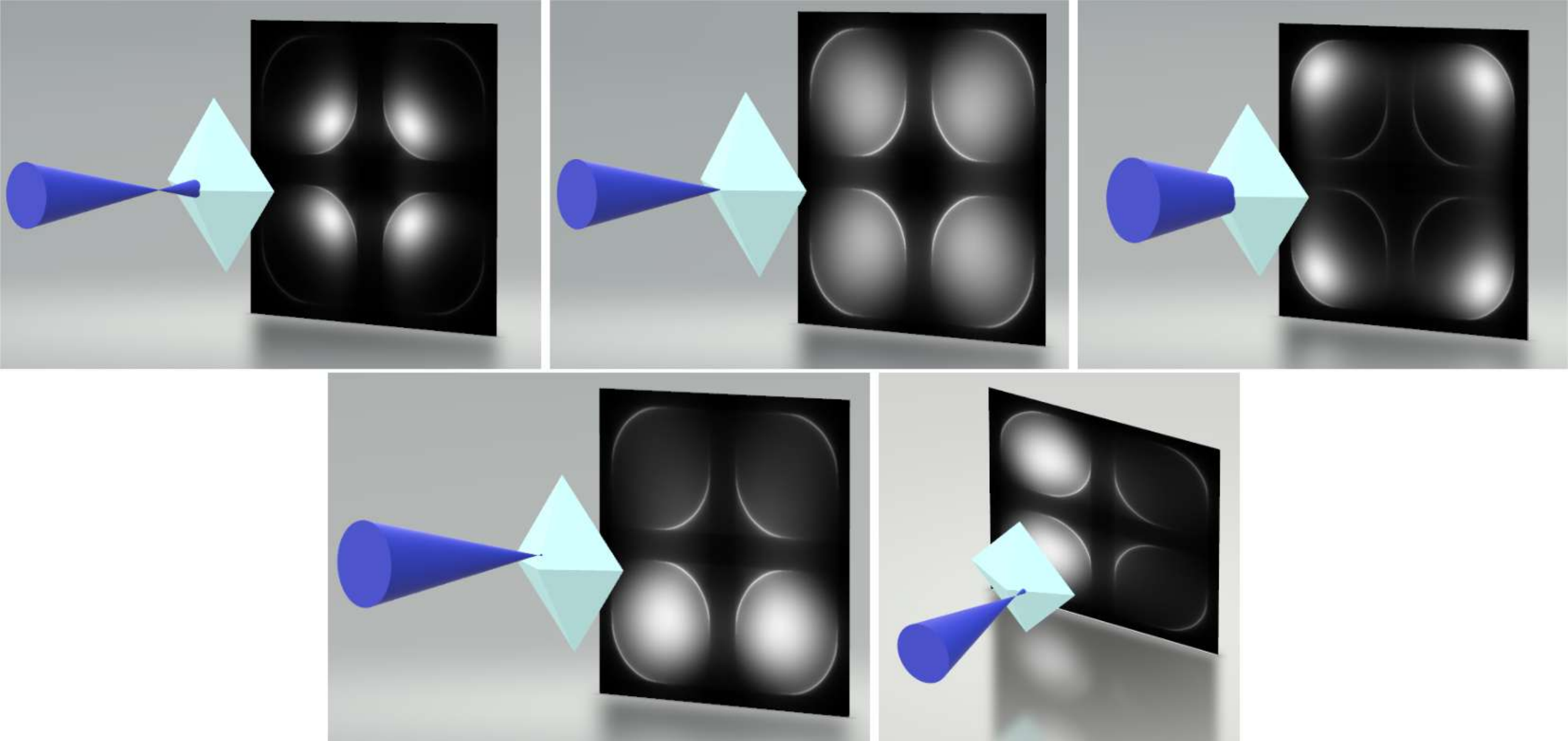}

    \caption{Illustration of the effects of positive, zero and negative focus (top row) and tip and tilt (bottom row) on the light distribution in the detector of the PWFS. The light cone is represented by the color blue and the glass pyramid by the cyan object.}
    \label{fig:ttf_pyr}
\end{figure*}

A simple schematic of the LGS and how its image interacts with the PWFS can be observed in Fig. \ref{fig:LGS_schematic}. The focal plane image of the LGS is not a perfect point source but rather has a width of around 1 arcsec and can be elongated in the order of tens of arcsec in one axis for a 40 m telescope due to the thickness of the sodium layer and the laser being launched from the side of the primary mirror. To understand the effects of this elongation, Fig. \ref{fig:ttf_pyr} shows how tip, tilt, and positive and negative focus impact the distribution of light in the detector of the PWFS. For each portion of the elongated LGS, the light distribution on the sensor will be a combination of these effects.

\begin{figure*}[t]
    \centering
    \includegraphics[width = \textwidth]{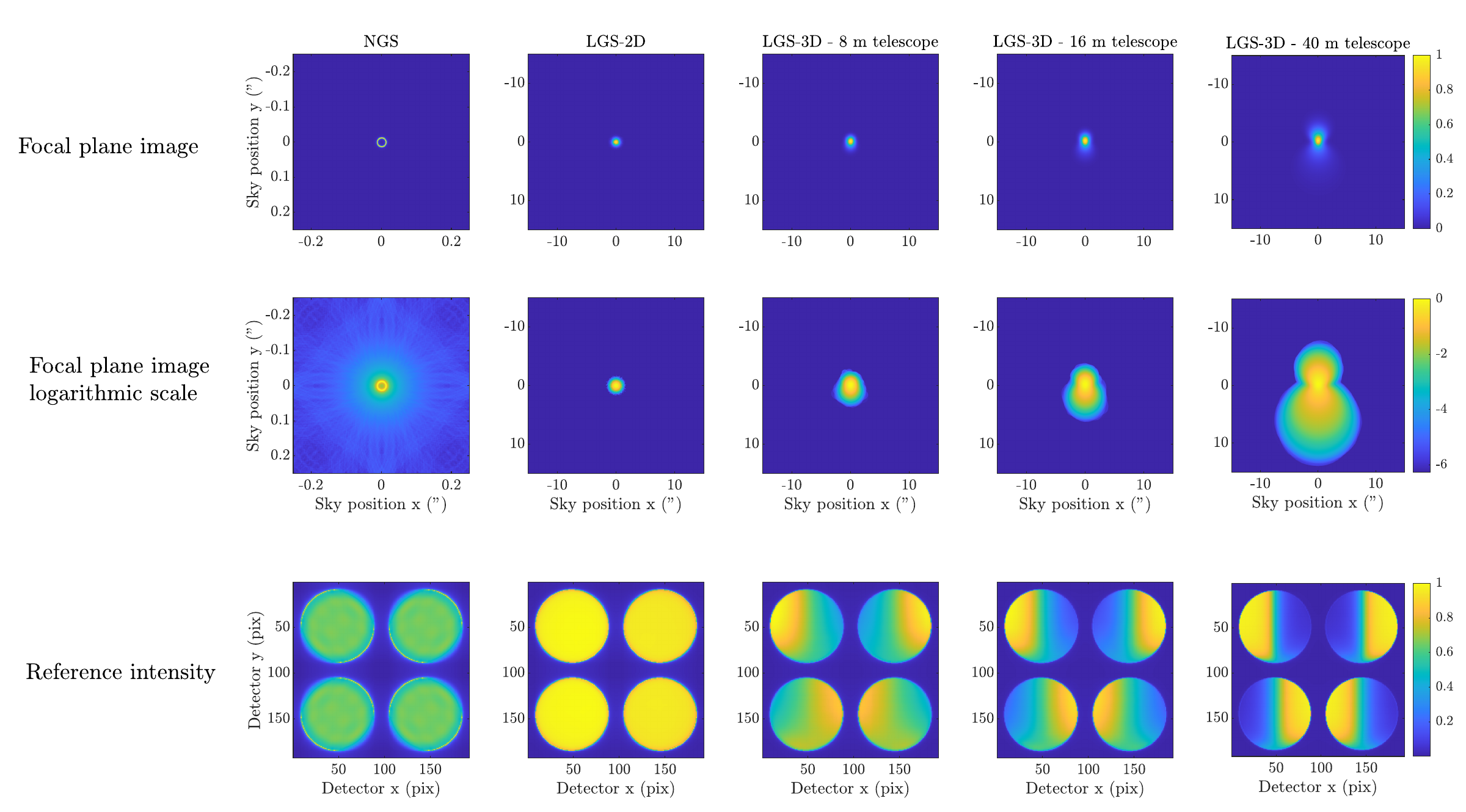}
    \caption{From left to right, the simulations correspond to a $4 \lambda/D$ modulated NGS for a 40 m telescope, LGS-2D, and LGS-3D for 8, 16 and 40 m telescopes. Top row: focal plane images. The scale was normalized such that the maximum pixel value in each image would be one; middle row: focal plane images with a logarithmic stretch for better visualization; bottom row: Intensities in the detector of the wavefront sensor for a flat wavefront. The intensities are normalized such that the maximum pixel value is one. Note that the field of view for the NGS corresponds to 0.4 arcsec, meanwhile, for the extended sources, the field of view is 30 arcsec. The last column includes a color bar that is valid for the whole row.}
    \label{fig:40_m_lgs}
\end{figure*}

Focal plane images of the test sources can be observed in the top and middle row of Fig. \ref{fig:40_m_lgs}. The first column from left to right is a $4 \, \lambda/D$ modulated NGS, the next column corresponds to a reference object composed of a non-elongated 1 arcsec spot. The following columns show the LGS focal plane image and detector intensity for progressively larger telescopes, from 8 up to 40 m in diameter.

To understand the light distribution on the detector of the PWFS shown in the bottom row of Fig. \ref{fig:40_m_lgs}, it is easier to divide the LGS into two halves. The top half is focused before the tip of the pyramid, skewing the light in the detector inwards, but due to the elongation, this portion mainly interacts with the upper faces of the pyramid, therefore the light gets refracted to the bottom pupils. Similarly, for the bottom half the light is focused after the tip of the pyramid and shifted toward the lower faces, making the light distribution on the detector skew outwards and to the top pupils. This effect is dependent on the telescope size, being more important for bigger telescopes, as for the 40 m telescope there is almost half of the pixels with little to no illumination.

An interesting case is to reduce the thickness of the sodium layer to zero, essentially obtaining an artificial guide star with no thickness, whose size would be determined by the width of the sodium laser and the atmospheric conditions, as can be observed in the left row in Fig. \ref{fig:40_m_lgs}. This case is interesting as it allows us to observe which elongation (x/y or along z) has a bigger impact on the PWFS sensitivity. Also, it allows us to obtain an upper bound on the performance of the system regarding noise propagation. An instrument can be built considering the Z elongation of the LGS, interacting with it as if it were only a 2D object. This case will be denoted LGS-2D, and when using the full 3D structure of the laser beacon it will be called LGS-3D. 

One interesting aspect to consider is the dependence of the elongation of the LGS on the Zenith angle. Both the angular size $\Delta \alpha$ and the extension normal to the focal plane $\Delta z$ are proportional to the cosine of the Zenith angle (see App. \ref{app:LGS_geom}), which means that the worst-case scenario for the LGS-3D corresponds when observing directly up. The more the zenith angle increases, the more similar the LGS-3D is to the LGS-2D.

It is important to consider that the size of the LGS acts similarly to the modulation with an NGS, as each sodium atom that emits light acts as a point source, and their contribution to the pyramid signal is incoherent with every other atom. One difference with the NGS, which usually is operated with modulation of a few $\lambda/D$ \citep{2020arXiv200307228S}, is the magnitude of this equivalent modulation, as 1 arcsec is equivalent to 65 $\lambda/D$ for an 8 m telescope, and around 330 $\lambda/D$ for a 40 m, considering for both cases $\lambda = 589 \, nm$. This equivalent modulation is responsible for lowering the sensitivity of the instrument \citep{2013aoel.confE..15Q}.

\subsection{LGS sampling}
To simulate a LGS it is necessary to discretize the sodium layer into samples. These samples correspond to individual point sources propagated through the PWFS, so that the PWFS signal is the incoherent sum of the signal produced by each sample. Previously, a common approach was to uniformly sample the LGS \citep{2010SPIE.7736E..57L, 2013aoel.confE..15Q, 2015aoel.confE..37B, 2016SPIE.9909E..6BE, 2018SPIE10703E..0VV}, dividing the sodium layer into regularly spaced slices, each of which containing regularly spaced point sources. Then, using a sodium density profile, the contribution of each layer was scaled to take into account the relative distribution of sodium atoms. There are several issues with this method of simulating an LGS:
(i) Many points have little contribution to the signal but are equally expensive computationally.
(ii) Large portions of the LGS are not sampled, therefore it is difficult to test real-like sodium profiles
(iii) The periodicity of the samples can introduce unwanted structures given by the symmetry and periodicity of the grid used for sampling.

\begin{figure*}
    \centering
    \includegraphics[width = \textwidth]{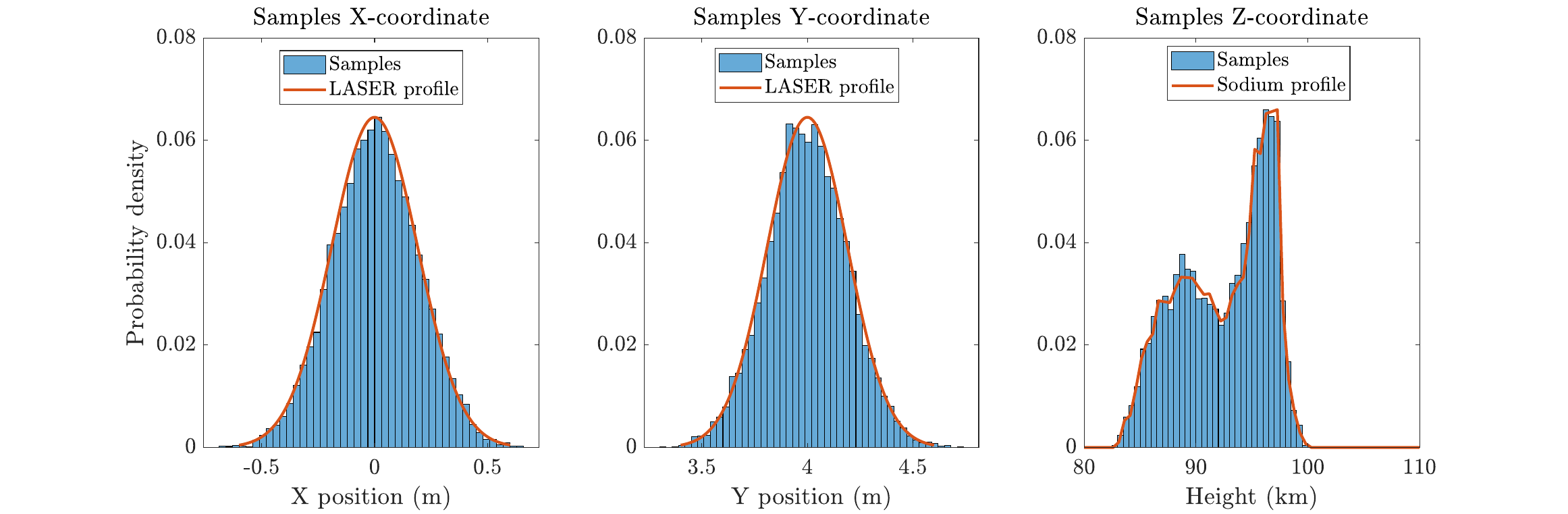}
    \caption{Probability density functions and sample histograms for the coordinates of each sample. The X and Y coordinates are randomly drawn from Gaussian distributions, whose width depends on the width of the laser beam, and the center of the Y distribution is on the edge of an 8 m telescope. The Z coordinate is randomly drawn from a probability density function that follows the sodium profile.}
    \label{fig:LASER_sodium_profiles}
\end{figure*}

\begin{figure}
    \centering
    \includegraphics[width = 0.49\textwidth]{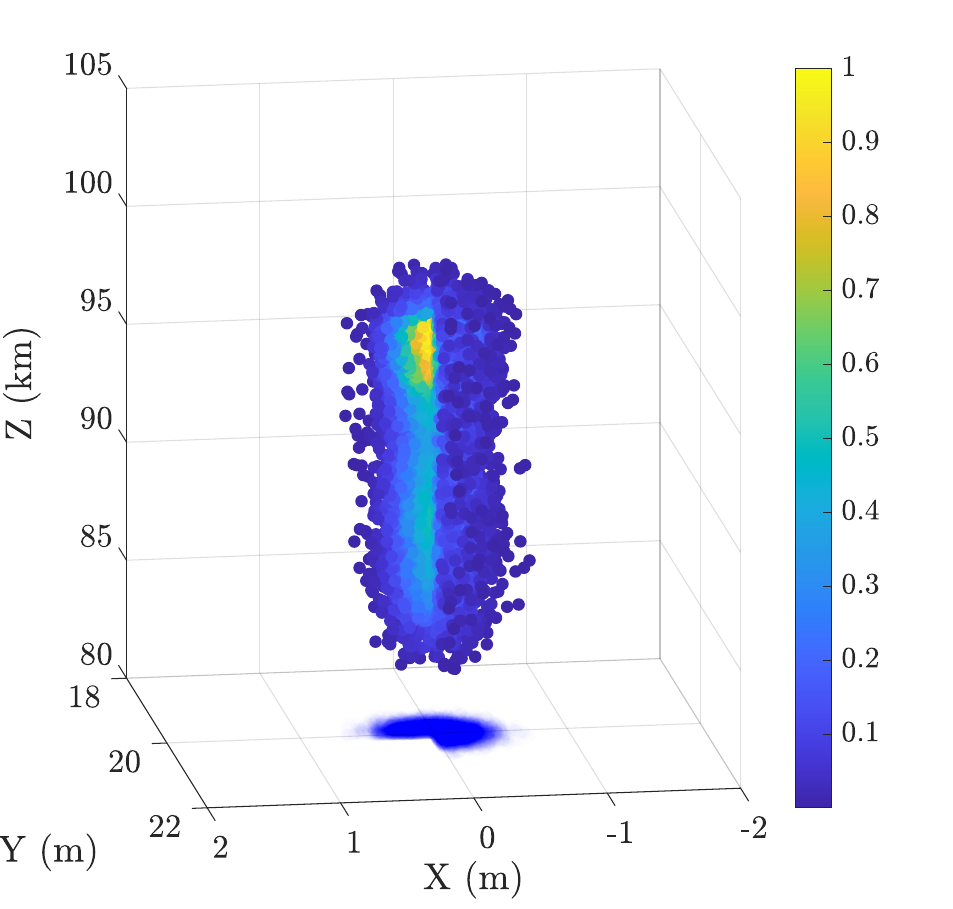}\\
    \includegraphics[width = 0.49\textwidth]{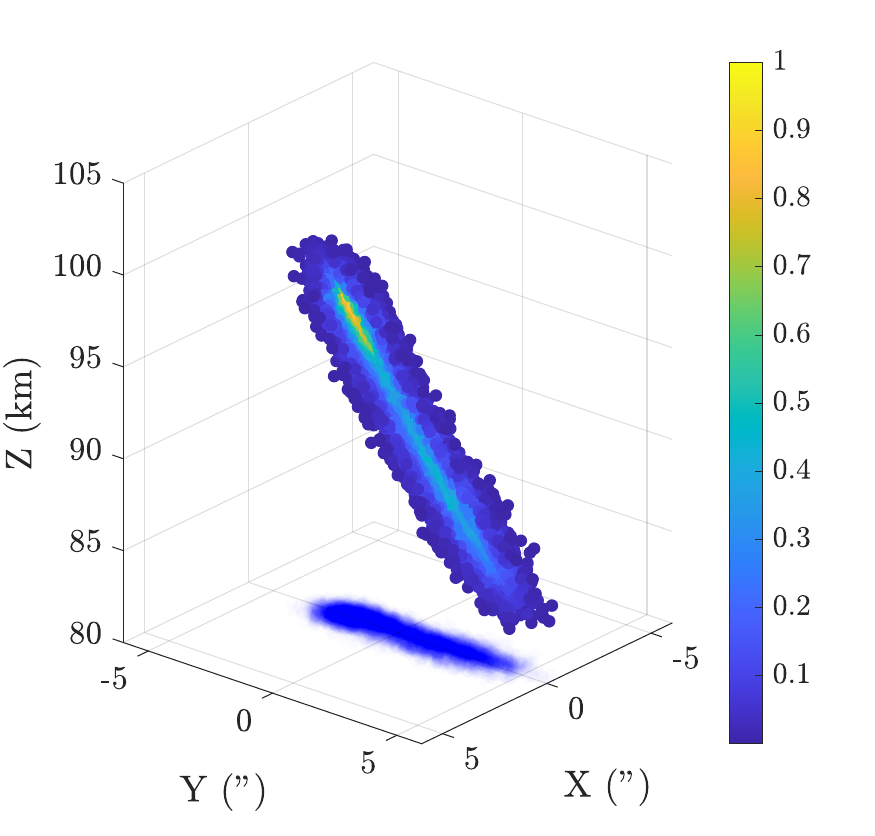}
    \caption{Top image: Position of each sample of the LGS in the sky for a 40 m telescope with the laser pointing parallel to the pointing of the telescope; bottom image: relative angular position of each sample with respect to the center of mass of the set of samples. The color of each sample corresponds to its probability density with respect to the most probable sample. A corner from both images was cut to show the inside of the LGS. The shadow on the bottom is there to show how the corner was cut from the LGS to show the inside and to show the dimensions of the source.}
    \label{fig:samples}
\end{figure}

Instead, in this work a Monte Carlo approach was used to simulate the LGS. To do this, the three coordinates of each sample were randomly drawn from sets that followed a specific probability density function. The X coordinate was drawn from a set that followed a Gaussian distribution centered at zero, with a Full Width at Half Maximum (FWHM) of the equivalent of 1 arcsec at 90 km, as can be observed in the left plot in Fig. \ref{fig:LASER_sodium_profiles}. The Y coordinate had the same FWHM but was centered at the side of the telescope, to simulate a laser being launched from the side of the primary mirror, as can be observed on the middle plot in Fig. \ref{fig:LASER_sodium_profiles}. For the Z coordinate, the relative distribution of the sodium atoms can be used as a probability density function, and generate a random set of samples that follows that distribution, as can be observed on the right plot in Fig. \ref{fig:LASER_sodium_profiles}. Fig. \ref{fig:samples} shows an example of the complete sampling of an LGS, where the color of each sample represents the relative density of samples, normalized such that the greatest probability is 1. The top image shows the physical location of each sample in the sodium layer, with a corner cut out such that is possible to observe the structure on the inside of the LGS. The bottom image shows the relative angular position of each sample as observed by the telescope (translated such that the center of mass is at the center), where it is possible to observe the elongation of the LGS.

Throughout this work, seven different sodium profiles are tested when possible, given the extensive simulation times needed to test each one. These sodium profiles correspond to typical conditions, each of which presents a distinctive characteristic, as presented in Fig. 2 of \citet{2014A&A...565A.102P}.

\subsection{A new, faster technique to simulate an LGS: ROI Propagation}

Building an interaction matrix can be a lengthy process, as we have to simulate two complete end-to-end propagations for each mode, taking into account the signal of each and every sample. If we consider the resolution of the telescope, the field of view and the number of corrected modes, it is possible to compute that the time to build an interaction matrix grows as the sixth power of the diameter.

With this in mind, we developed a new technique we call \textit{Region of Interest (ROI) Propagation} that tackles the problem of the large field of view needed to simulate the LGS. It is based on the fact that each individual sample of the LGS interacts with a small portion of the pyramid \citep{2022SPIE12185E..4WO}. The general idea of the method is instead of using the full field of view needed to accommodate the LGS, we take only the portion of the pyramid the sample would interact with. To do this, we translate and crop the pyramid phase mask to compensate for the tip and tilt of the sample. To simulate the LGS-3D, we add the focus coefficient of each sample, given by its distance from the telescope. For the 2D version, this step is skipped. Finally, we propagate each sample individually and incoherently sum the PWFS signals of all the LGS samples to obtain the final signal.

To explain the mathematical basis of this method, let's consider the incoming wavefront of a single sample. This wavefront will have contributions to its phase coming from the atmosphere $\phi_{atm}$, the height of the sample $\phi_{focus}$, and its position in the sky $\phi_{til-tilt}$ with respect to the pointing of the telescope. At first, let's only consider the contributions of the atmosphere and the height of the sample. The wavefront can be expressed as

\begin{equation}
    \psi(x,y) = \mathbb{I}_p(x,y) e^{i(\phi_{atm}(x,y) + \phi_{focus}(x,y))},
\end{equation}

with $x$ and $y$ the coordinates of the entrance pupil and $\mathbb{I}_p$ the pupil indicative function. A simplified form of the wavefront at the focal plane $\Psi$ can be obtained by taking the Fourier transform of the wavefront at the pupil plane

\begin{equation}
    \Psi(u,v) = \left.\mathcal{F}(\psi(x,y))\right|_{u = \frac{x}{\lambda}, v = \frac{y}{\lambda}}.
\end{equation}

Now, considering that the sample is located at an angle displacement of $(\alpha, \beta)$ with respect to the pointing of the telescope, the wavefront at the focal plane is displaced by that amount, 

\begin{equation}
    \Psi(u-\alpha,v-\beta) = \left.\mathcal{F}(\psi(x,y) \, e^{i\phi_{tip-tilt}(x,y)})\right|_{u = \frac{x}{\lambda}, v = \frac{y}{\lambda}}.
\end{equation}

This wavefront is then affected by the phase mask $m$ characterized by 

\begin{equation}
    m(u,v) = e^{i\phi_{mask}(u,v)},
\end{equation}

and then, considering $x'$ and $y'$ the coordinates of the detector plane, the light distribution of the pupil images $I(x',y')$ is obtained by taking the Fourier transform of the combination of the wavefront at the focal plane and the phase of the mask

\begin{equation}
    I(x',y') = \left|\left.\mathcal{F}(\Psi(u-\alpha,v-\beta) \, m(u,v))\right|_{x = \lambda u, y = \lambda v}\right|^2.
\end{equation}

Using the properties of the Fourier transform, the previous expression can be expressed as 

\begin{equation}
    I(x',y') = \left|\left.\mathcal{F}(\Psi(u,v) \, m(u+\alpha,v+\beta))\right|_{x = \lambda u, y = \lambda v}\right|^2.
\end{equation}

As we get back $\Psi(u,v)$, it means that instead of having to simulate the complete field of view for each sample, the phase mask can be shifted to take into account the position of each sample with respect to the pointing of the telescope, and the field of view can be adapted for each sample such that it is big enough to contain the complete image of the star.

\begin{figure}
    \centering
    \includegraphics[width = 0.49\textwidth]{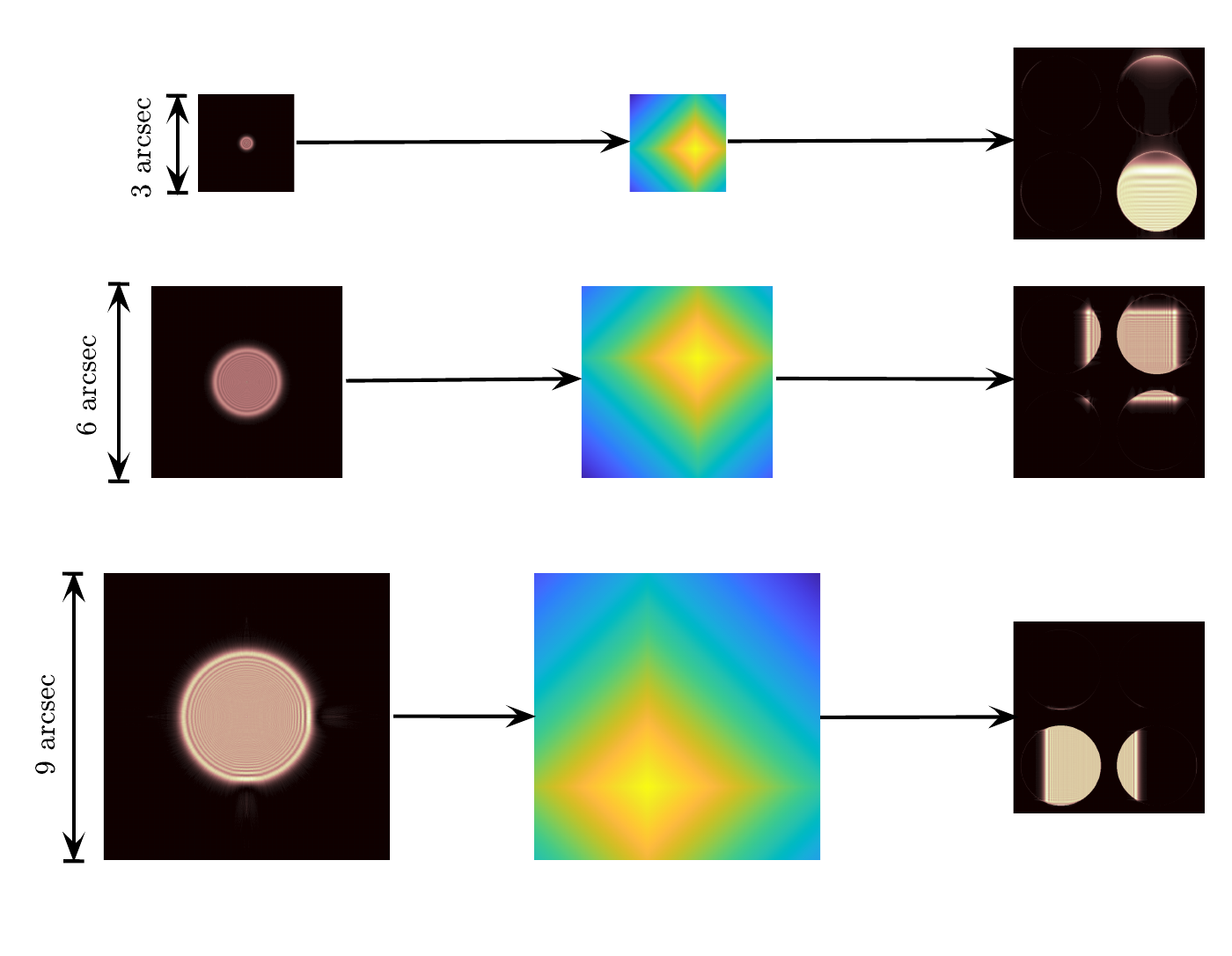}
    \caption{Example for a flat wavefront of three samples propagated using the ROI Propagation technique for a 40 m telescope. From top to bottom, the samples are at 400, 3000 and 6000 meters away from the plane of focus.}
    \label{fig:portion_diagram}
\end{figure}

To illustrate how this works, Fig. \ref{fig:portion_diagram} shows an example for a 40 m telescope of three samples at 400 (top), 3000 (middle) and 6000 (bottom) meters away from the plane of focus, located at 90 km from the telescope, where the left column shows the image of the sample in the focal plane, the middle column shows the region of interest of the pyramidal mask the sample is interacting with and on the right column the image of the pupils in the detector plane.

\begin{figure}
    \centering
    \includegraphics[width = 0.49\textwidth]{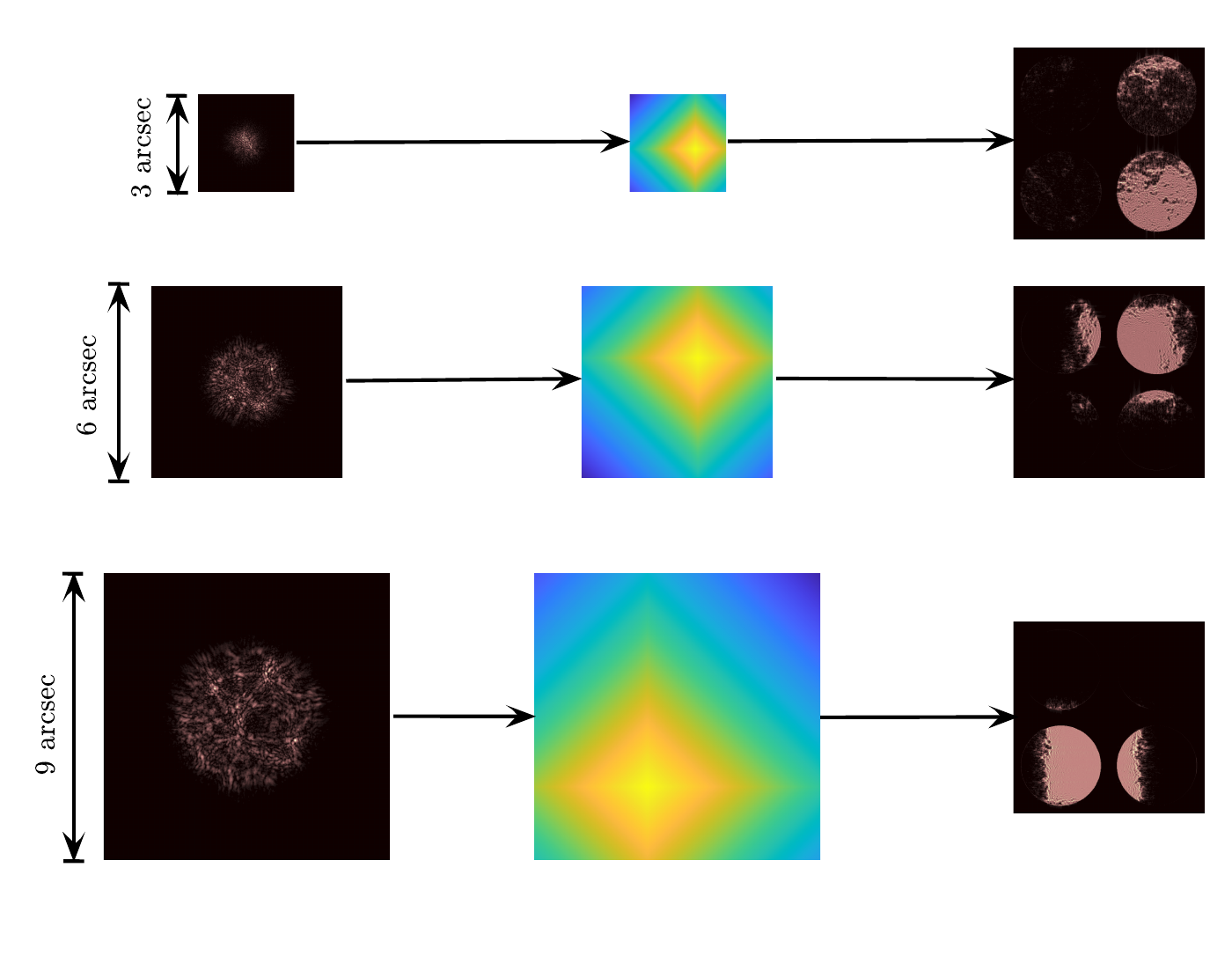}
    \caption{Three examples of seeing limited samples propagated using the ROI Propagation technique for a 30 m telescope. From top to bottom, the samples are 400, 3000 and 6000 meters away from the plane of focus.}
    \label{fig:portion_diagram_atm}
\end{figure}

Special care has to be taken when including atmospheric turbulence, to contain the complete image of the sample within the selected field of view. Fig. \ref{fig:portion_diagram_atm} shows how this method handles seeing limited samples. 

\begin{figure}
    \centering
    \includegraphics[width = 0.49\textwidth]{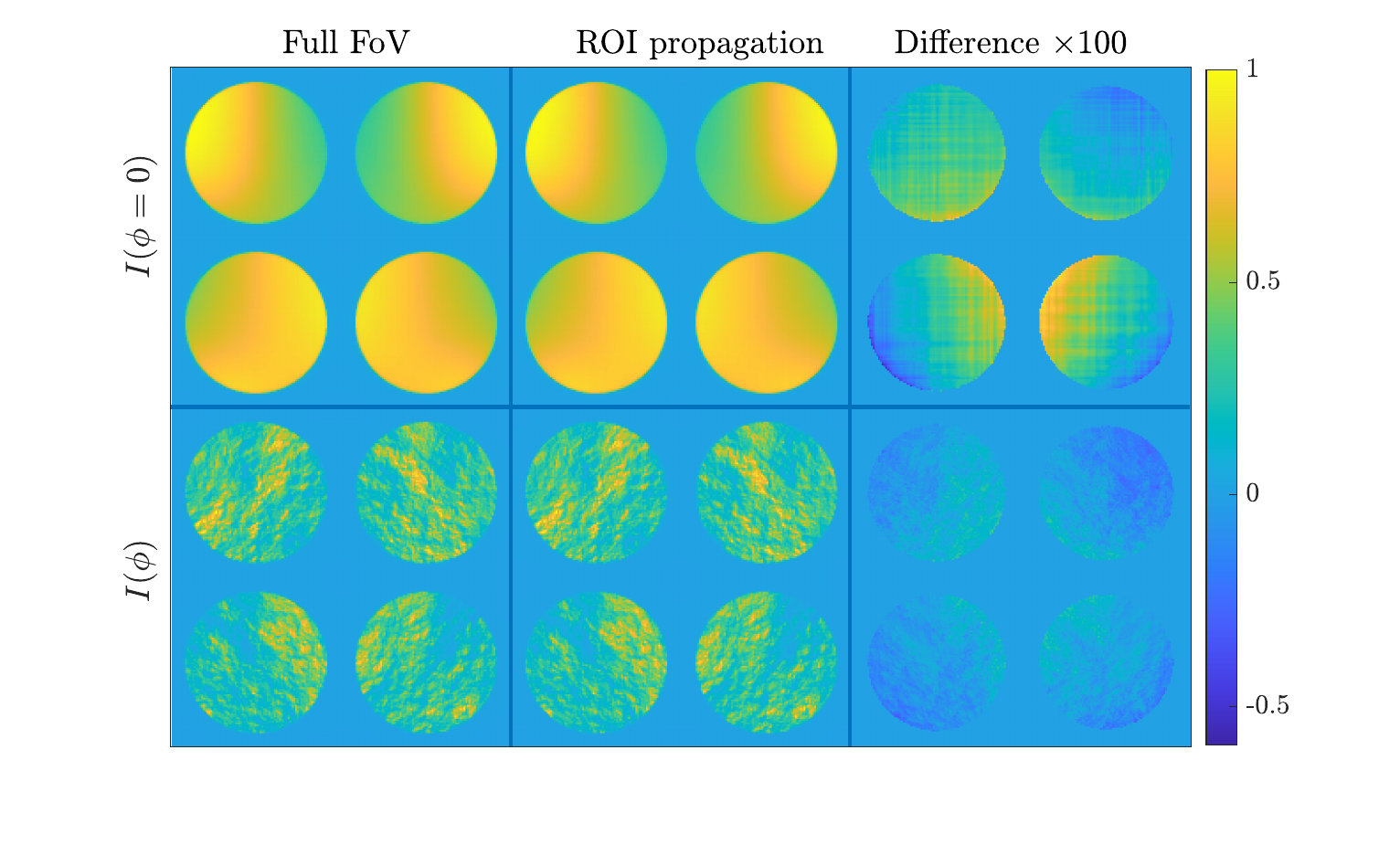}
    %\caption{Left column: Full E2E simulation considering the complete FoV for each sample; middle column: Portion propagation; right column: the difference between the two methods. Top row: Intensity for a flat wavefront; middle row: intensity considering atmosphere; bottom row: Reduced intensities considering atmosphere}
    \caption{Left column: Full simulation considering the complete FoV for each sample; middle column: ROI Propagation; right column: 100 times the difference between the two. Top row: Intensity for a flat wavefront; bottom row: intensity considering atmosphere}
    \label{fig:E2E_portion}
\end{figure}

To check if this new method produced the same signal as the complete E2E considering the full FoV for each propagation, we simulated both a propagation for a flat wavefront and another with atmospheric conditions for an 8 m telescope. We then normalized the images and computed the difference, finding that for each case both methods produce practically the same signal. Fig. \ref{fig:E2E_portion} shows the images obtained in each simulation, where the right column corresponds to the difference between the two methods. To give a metric of how similar these methods are, we took the RMS of the difference of the atmosphere-affected pupils $I(\phi$) and we found an RMS value less than 0.1 \% of the average pixel value. For this simulation, the new method was computed over 30 times faster. As the telescope size gets bigger, this new method provides even greater speedup with respect to the full FoV method.

It is important to remark that even with this new method, a full end-to-end interaction matrix for the 40 m telescope is too demanding both in time and computational resources: we could not perform with our current hardware, as it would have taken months to compute. Nevertheless, in the next section, we used it to speed up the E2E simulations of smaller telescopes up to 24 m in diameter, to explore an alternative calibration procedure that would allow us to obtain an approximation of the interaction matrix for the 40 m telescope, by using one calibrated using a point source.

\section{Interaction matrix for a laser guide star}
\label{sec:IMat_cal}

\subsection{Optical gains}

When calibrating the interaction matrix for a real instrument, it can be challenging to use an LGS-like source. Therefore, it is necessary to test an ideal case where we can calibrate on an LGS and a more realistic case where we use a point source for calibration.
To be able to compare the signal obtained from these two calibration procedures, we can use the interaction matrix of each to observe differences in signal intensity and composition. To compare two signals, a reference signal $a$, and a test signal $b$, the interaction matrix calibrated using source $a$ is used as the reconstructor and the interaction matrix calibrated using source $b$ as the signal (recall equation \ref{eq:fst_recon}, but instead of a vector of measurements $\Delta I(\phi)$ we use the complete interaction matrix calibrated using source $b$). Using this we obtain what we call a Modal Transform Matrix (MTM). Mathematically, 

\begin{equation}
\begin{split}
&MTM_{a\rightarrow b} = \mathcal{D}^\dagger_{a} D_{b}
\\
&OG_{a\rightarrow b} = diag(MTM_{a\rightarrow b}),
\end{split} 
    \label{eq:MTM}
\end{equation}

where the diagonal of the MTM, known as optical gains (OG) \citep{2008ApOpt..47...79K, 2018SPIE10703E..20D, 2021A&A...649A..70C}, corresponds to the intensity of the signal obtained using $b$ as a guide star when compared to $a$, and the non-diagonal terms correspond to mode confusion (i.e. the difference in the structure of the signal). As an example, if $MTM_{a\rightarrow b}(i,i) = 0.7$, that means you lose $30 \% $ of the signal intensity for mode $\phi_i$ if you change from source $a$ to $b$.

\begin{figure}
    \centering
    \includegraphics[width = 0.45\textwidth]{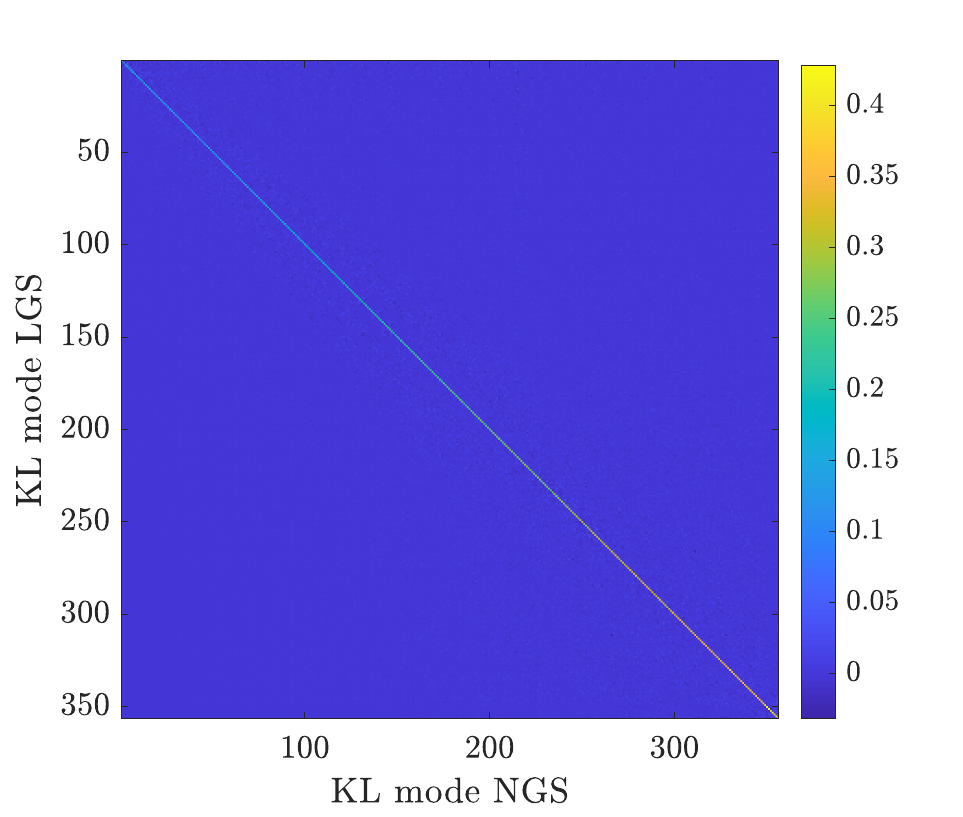}
    \caption{Simulation of the modal transfer function $MTM_{NGS \rightarrow LGS}$ for an LGS with an 8 m telescope using an NGS with $4 \, \lambda/D$ modulation as reference. The values of the diagonal encode the optical gains. The matrix is mainly diagonal with few non-diagonal terms.}
    \label{fig:OG_8m}
\end{figure}

Using the ROI Propagation technique, we computed the end-to-end interaction matrix $\mathcal D_{LGS}$ for the first 350 KL modes for 8, 16 and 24 m telescopes. We used only 350 modes for this test, given that for the 24 m telescope this process took days to compute. Using an NGS with 4 $\lambda/D$ modulation as a reference, we computed the $MTM_{NGS \rightarrow LGS}$ and for each telescope diameter we got a matrix that was mainly diagonal, as it can be seen in Fig. \ref{fig:OG_8m}, where we show the example of the 8 m telescope. This diagonal structure means that the signal coming from an LGS has almost the same structure as the NGS, but is attenuated by the value in the diagonal.

This attenuation of the signal comes from the spreading of the light of the LGS over a larger area of the pyramid than the NGS. The pyramid mainly produces a signal from the light that interacts with its edges and for the LGS, a large portion of the light falls in the faces of the pyramid, which from the point of view of the rays of light is just an inclined plane of glass. As this produces no filtering, there is no signal.

The fact that the signal from the LGS is closely related to the NGS's means that it is possible to build the interaction matrix using a point source as a reference, and then optimize it for the LGS by multiplying it by a diagonal matrix whose elements are the optical gains $OG_{NGS \rightarrow LGS}$.

\begin{equation}
    \mathcal{D}_{LGS} \approx \mathcal{D}_{NGS} \, OG_{NGS \rightarrow LGS}.
    \label{eq:D_NGS_LGS}
\end{equation}

This not only gives an interesting alternative to compute the interaction matrix for the LGS in simulation, but it also has implications in a real telescope, as the calibration source would most likely be a point source. The interaction matrix for the point source $\mathcal{D}_{NGS}$ is relatively easy to obtain in a real scenario, but the optical gains needed to optimize the reconstructor for the LGS might not be, as they depend on the ever-evolving structure of the sodium layer, making the pre-computation of these values not too effective.

One way to have access to an approximate interaction matrix for the LGS with a real telescope, and therefore to the optical gains, would be with the introduction of a gain scheduling camera and the use of a convolutional model, as proposed in \citet{2021A&A...649A..70C}, where the focal plane image of the guide star is used to estimate it. In the next section, we show that it is possible to use this technique with an extended object such as an LGS to obtain the optical gains. Using this method it is also possible to keep track of the evolving structure of the sodium layer, optimizing the reconstructor to accommodate the changes in the LGS.

\section{Accessing the optical gains by means of a convolutional model}
\label{sec:CMod}

The convolutional model introduced by \citet{2019JOSAA..36.1241F} can predict the signal of a Fourier filtering wavefront sensor by means of computing its impulse response (IR) (signal when introducing Dirac's delta in phase). This model makes assumptions on linearity, which are a simplification of complete E2E propagation. This makes its predictions less accurate, but much more efficient. If we have an FFWFS characterized by a mask $m$ and the focal plane image of the source $\omega$ interacting with the mask (Fig. \ref{fig:40_m_lgs} top), the IR can be computed as

\begin{equation}
    \label{eq:IR}
    IR = 2 \, \textbf{Im} \left[\, \overline{\hat m} \hat{(m \omega)} \right],
\end{equation}

where $^-$ is the conjugate operator and $\hat{\mkern6mu}$ the Fourier transform. Using this tool, the reduced intensity of a FFWFS can be obtained as 

\begin{equation}
    \Delta I(\phi) = (\mathbb{I}_p\phi) \star IR,
\end{equation}

where $\star$ denotes the convolution operation and $\mathbb{I}_p$ the pupil indicative function. To be able to use this tool, a single focal plane image of the LGS has to be computed to get the impulse response, and with that, each column of the interaction matrix can be obtained by a single convolution, reducing the time needed to compute it almost $2N$ times, where $N$ is the number of samples used.

\begin{figure}
    \centering
    \includegraphics[width = 0.49\textwidth]{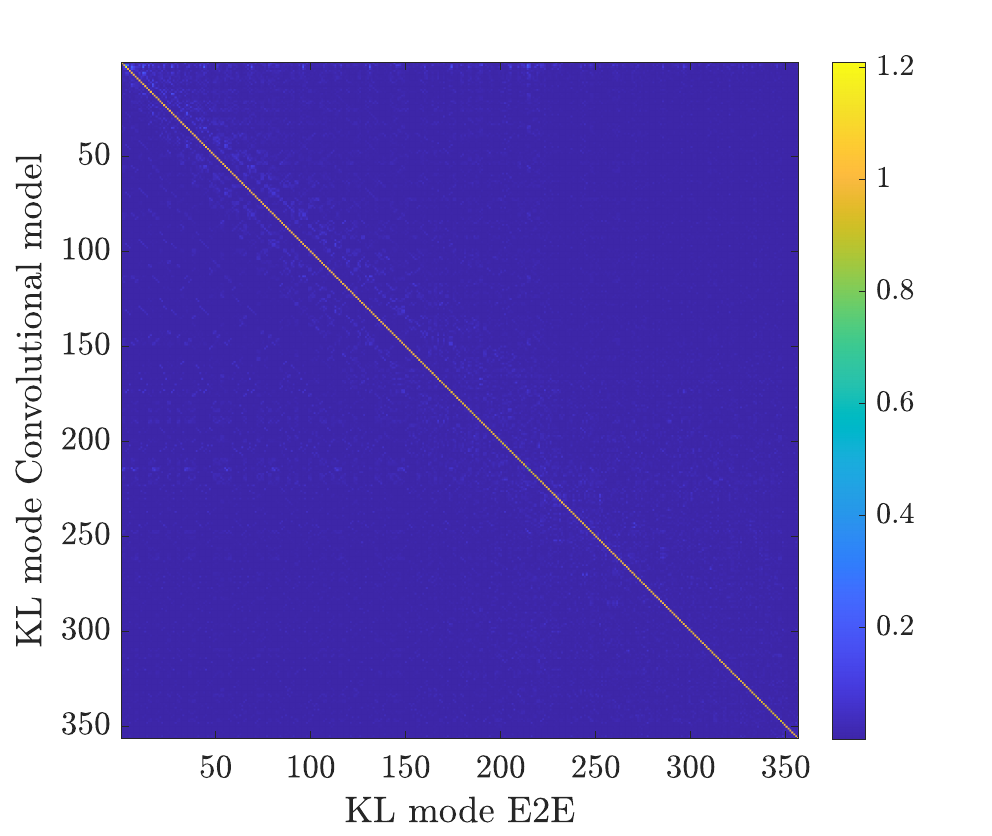}
    \caption{$MTM_{LGS \rightarrow Conv}$ matrix obtained using equation \ref{eq:MTM} when simulating a LGS for a 16 m telescope.}
    \label{fig:OTM_16m}
\end{figure}

As the convolutional model has mainly been tested with 2D modulated NGS, we first had to test the validity of using this model with an extended 3D object, such as an LGS. To do this, we used the convolutional model to re-compute the interaction matrices $\mathcal D_{Conv}$ for the 8, 16, and 24 m telescopes we had previously computed using end-to-end methods. We then compared them by computing the Modal Transfer Matrix. The $MTM_{LGS \rightarrow Conv}$ for each telescope diameter was similar to the identity, meaning that the convolutional model accurately predicts the structure and intensity of the signal produced by the LGS. Fig. \ref{fig:OTM_16m} shows the $MTM_{LGS \rightarrow Conv}$ for the 16 m telescope.

\begin{figure}
    \centering
    \includegraphics[width = 0.49\textwidth]{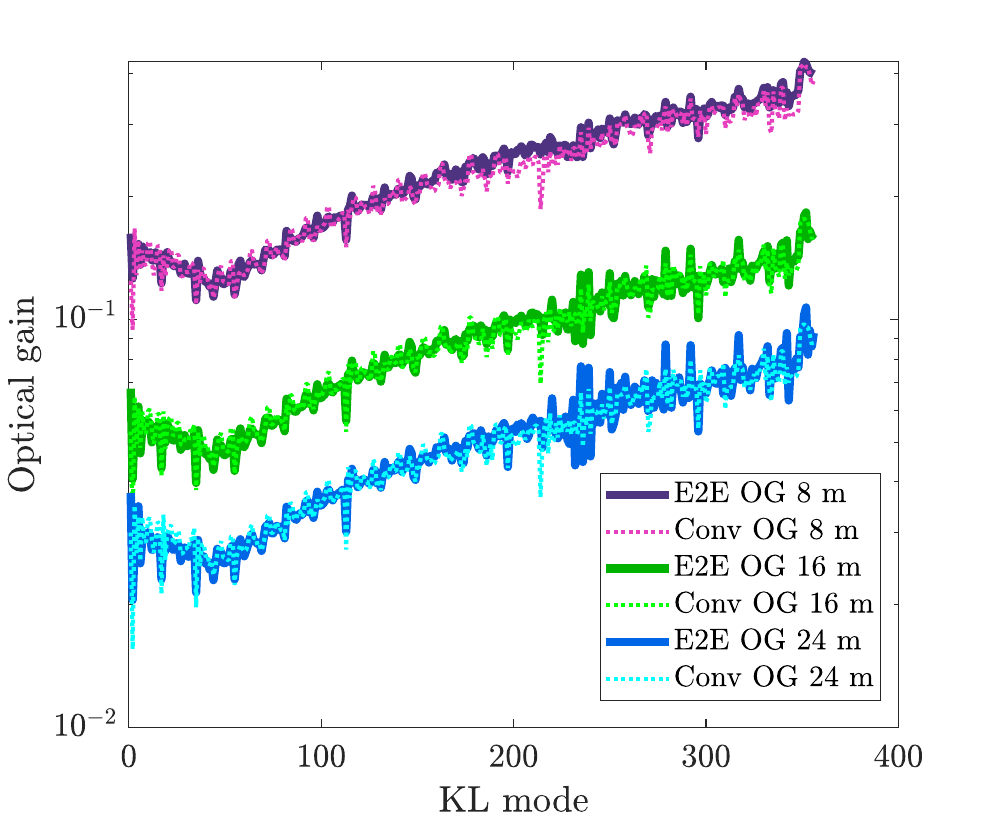}
    \caption{Optical gains using E2E (solid line) and convolutional model (dashed line) for 8, 16 and 24 m telescopes.}
    \label{fig:OG_LGS_e2e_conv}
\end{figure}

Now, by using the convolutional model we can predict the values for the optical gains needed to optimize the reconstructor, and compare them to the optical gains computed using the full E2E methods, as seen in Fig. \ref{fig:OG_LGS_e2e_conv}. The optical gains predicted by the convolutional model closely match the ones obtained using the E2E simulation, meaning that the model can be used to compute the values needed to optimize the reconstructor such that it can properly use the signal of the LGS.

From Fig. \ref{fig:OG_LGS_e2e_conv} it is possible to observe that the convolutional model struggles to estimate correctly the optical gains at the first few modes, corresponding to low spatial frequency. This could have a large impact on Tip, Tilt and Focus, but as we are dealing with an LGS, we won't be measuring those modes. The model also has a problem recovering the value of the optical gains for specific modes across all telescope sizes (e.g. mode 211 is predicted with a lower value). These modes have their intensity localized more on the edges of the pupil. This causes problems with the convolutional model, as the discontinuity of the pupil indicative function is considered as signal, which is exaggerated for modes that have more energy at the edges.

We recommend using the convolutional model just to compute the values of the optical gains, and with them optimize the interaction matrix to the LGS, instead of using the interaction matrix obtained with the model directly. This is because the model has issues with the discontinuities in the pupil indicative function, predicting an excess of signal in the edges of the pupil images. This is especially noticeable for low spatial frequencies. If the interaction matrix obtained with the convolutional model is used, then it is probable that the corrections at the center of the pupil might behave properly, but a large accumulation of phase might occur at the edges. 

The optical gains change for each telescope size. This is because as the telescope size gets bigger the relative size of the LGS gets bigger with respect to a diffraction-limited spot, as 1 arcsec is equivalent to 65 $\lambda/D$ for an 8 m telescope and almost 200 $\lambda/D$ for a 24 m one. This is responsible for decreasing the amount of light of the LGS that is split by the edges of the PWFS, and it is mainly in the edges where the signal of the PWFS is coming from, lowering the strength of the signal. This results in a decrease in optical gain as the diameter of the telescope increases.

On a real telescope, introducing a gain-scheduling camera would allow obtaining focal plane images of the LGS, and then using the convolutional model it would be possible to compute the interaction matrix for the LGS. It can then be used to compute the optical gains needed to update the point source calibrated interaction matrix using equation \ref{eq:D_NGS_LGS}.

\begin{figure}
    \centering
    \includegraphics[width = 0.49\textwidth]{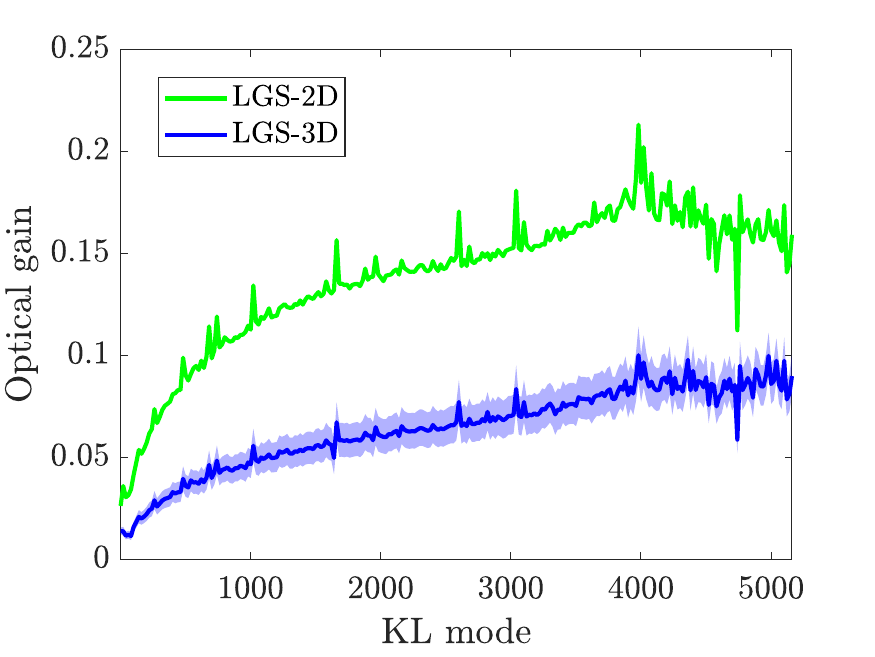}
    \caption{Optical gains for a 40 m telescope needed to optimize the reconstructor. In green are the optical gains for the LGS-2D, and in blue for the LGS-3D. The solid line corresponds to the average optical gain across all sodium profiles and the shaded region is limited by the maximum and minimum optical gain at each mode}
    \label{fig:OG_40m}
\end{figure}

With this tool, now it is possible to compute the optical gains needed to optimize the reconstructor calibrated with a point source to accommodate the signal of an LGS for a 40 m telescope. We computed the optical gains for 5100 modes for every sodium profile for the LGS-3D, and for the LGS-2D, and plotted them in Fig. \ref{fig:OG_40m}. It is possible to observe that the optical gains for the LGS-3D are smaller than for the LGS-2D, which is expected, given its larger size. On average, the optical gains for the LGS-3D are 2.5 times smaller, meaning that the equivalent size of the LGS-3D is about 2.5 times larger than the LGS-2D.

The values of the optical gains vary up to 40 \% when changing the sodium profile, which implies that it would be best if an online system is continuously updating these values, given that in the worst-case scenario the gain could be up to 40\% off. The frequency that the gain-scheduling camera should take images to update the optical gains should follow the time scale of the changes in the sodium layer, which typically is in the order of a few minutes \citep{2014A&A...565A.102P}.

Optical gains encode the strength of the signal when compared to the signal obtained with an NGS, which also implies a loss in sensitivity. These quantities, optical gain and sensitivity are proportional to each other, meaning that, for example, for a given mode $\phi_i$ an optical gain of 0.3 implies that the WFS has $30 \, \%$ of the sensitivity when using a NGS. Considering equation \ref{eq:sigma_gamma}, this means around 11 times more residual variance being introduced.

When going from an NGS to an LGS-2D, there is a reduction in the intensity of the signal of around 10-30 times for low-order modes (<1000) and around 5-10 times for higher-order. Then if we take into consideration the thickness of the sodium layer, i.e. going from the LGS-2D to the LGS-3D, then there is a drop in the intensity of the signal of about 2.5 times. This suggests that the biggest drawback on the PWFS sensitivity with extended objects is the width of the laser source, rather than the thickness of the sodium layer.

Now that we have shown that the convolutional model can be used to compute the optical gains, it only takes a fraction of the time to build the interaction matrices for extended objects. In the next section, we will use these to perform end-to-end closed-loop simulations of smaller telescopes, and then, in combination with the sensitivity model to predict the expected performance for the 40 m telescope.

\section{Performance of the AO loop}
\label{sec:closed_loop}

\subsection{End-to-end simulations}

We performed E2E simulations of a closed-loop control system for a $4\, \lambda/D$ modulated NGS, the LGS-2D and LGS-3D for 8 and 16 m telescopes. These simulations had two main purposes: first, we wanted to test if the point source calibration, optimized using the convolutional model would work in a closed loop for the extended objects, and second to test if the analytical model would predict the performance of the AO loop.

The simulation parameters are found in table \ref{tab:simulation_params}. For the throughput of the telescope, we chose to use 100 \%, as it will later be easier to adapt these results when estimations of the actual efficiency of the system are computed. For the zenith angle we chose 0 degrees because it will show the biggest difference between the LGS-2D and LGS-3D. Choosing a higher zenith angle would make the difference smaller, as discussed in Sec. \ref{sec:LGS}. To select the number of samples of the LGS, we simulated the sensitivity to photon noise for a 16 m telescope. We tested samples ranging from $100$ to $100,000$ and we found that beyond $10,000$ samples, the values of the sensitivities did not change, therefore we picked that amount for the rest of the simulations.

The cone effect will have a large impact on the residual phase when using a single LGS, and for that reason, the ELT will use multiple lasers for tomographic reconstruction of the wavefront \citep{2016SPIE.9908E..1XT, 2022SPIE12185E..14C}. But, as it is not correlated with photon noise, we chose to use a single ground layer to discard its effects, as it is outside the scope of this work. For the same reason, we chose to use a static atmosphere, because the moving atmosphere would interfere with the measurements of the impact of photon noise. The detectors for the PWFS we used had enough pixels such that aliasing would have a minimum impact. For this reason, the 8 m telescope had 60 x 60 subapertures and the 16 m telescope had 80 x 80 subapertures. As we tested the performance for different telescope sizes, we used a DM pitch of $50$ cm as a constant across all simulations. This meant that for the 8 m telescope, we used 17 actuators across the pupil and for the 16 m, 33. For the science wavelength, we chose to be the same as the sensing wavelength. The residual phase due to noise will depend on the science wavelength, but the differences between the NGS and the extended objects will remain the same.

\begin{table}
    \caption{Simulation parameters}
    \label{tab:simulation_params}
    \centering
    \begin{tabular}{l l}
    \hline
    \hline
    \textbf{Telescopes}  &  \\
    Diameter     &        8.0, 16.0\\
    Throughput & 100 \%\\
    Central obstruction & None\\
    \\
    \textbf{Natural guide star}  &  \\
    Zenith angle & $0^o$\\
    Magnitudes & 5-19\\
    Zero point & $8.96 \times 10^9$ $photons/s/m^2$\\
    Modulation & $4 \, \lambda/D$\\
    \\
    
    \textbf{Laser guide star} & \\
    Zenith angle & $0^o$\\
    Magnitudes & 5-14\\
    Zero point & $8.96 \times 10^9$ $photons/s/m^2$\\
    Number of samples & 10,000\\
    Sodium profile &  TopHatPeak \\
    \\
    \textbf{Atmosphere} & \\
    $r_0$ & 15 cm \\
    $L_0$ & 25 m\\ 
    Layers & 1\\
    Altitudes & 0 m\\
    Wind speed & 0 m/s\\
    \\
    \textbf{WFS} & \\
    Order for 8 m telescope & 60 x 60 subapertures\\
    Order for 16 m telescope & 80 x 80 subapertures\\
    Frequency & 1 KHz\\
    $\lambda_{sens}$ & 589 nm\\
    \\
    \textbf{DM} & \\
    Order for 8 m telescope & 17 x 17 actuators (200 KL modes)\\
    
    Order for 16 m telescope & 33 x 33 actuators (800 KL modes)\\
    \\
    \textbf{AO loop} & \\
    Delay & 2 frames \\
    Gain & 0.3 \\
    \\
    \textbf{Science} & \\
    $\lambda_{sci}$ & 589 nm\\
    \hline
    \end{tabular}
\end{table}

The interaction matrices for the extended objects were built using the convolutional approach to compute the optical gains and then optimize an interaction matrix calibrated using a point source to accommodate the signal coming from the LGS, as described in Sec. \ref{sec:IMat_cal} and \ref{sec:CMod}. By iterating we found that it is best to use a high modulation ($> 20 \, \lambda/D$) for the point source calibrated interaction matrix, for the PWFS to be in a similar sensing regime as with the extended objects.

As a real LGS gives no information about the global tip or tilt and has problems with the focus term given the evolving structure of the sodium layer, we ran in parallel a noiseless AO loop that controlled tip, tilt and focus. This also implied that we did not consider the sensitivities for these three modes in the predictions of the total noise transmitted through the loop.

\begin{figure}
    \centering
    \includegraphics[width = 0.49\textwidth]{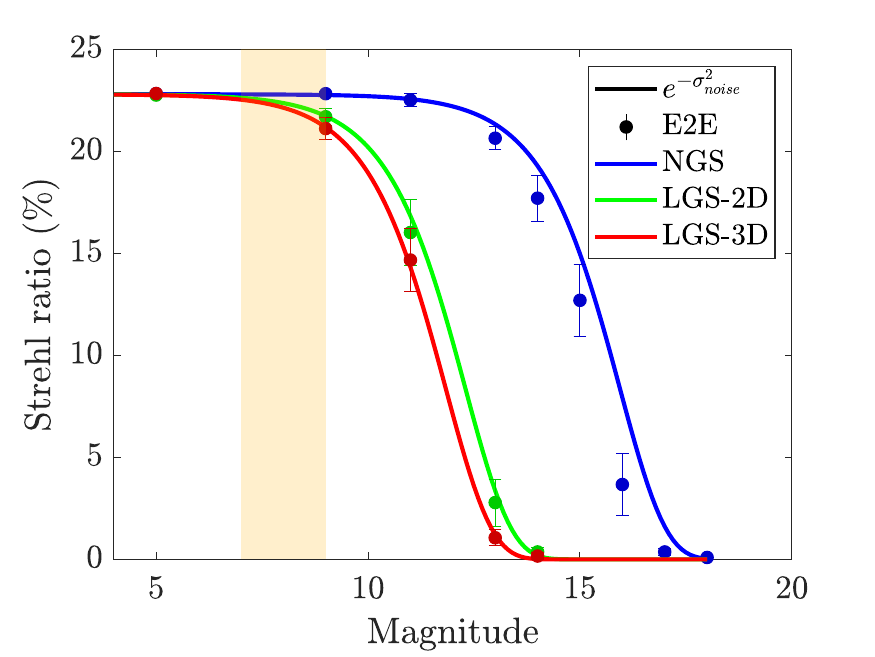}\\
    \includegraphics[width = 0.49\textwidth]{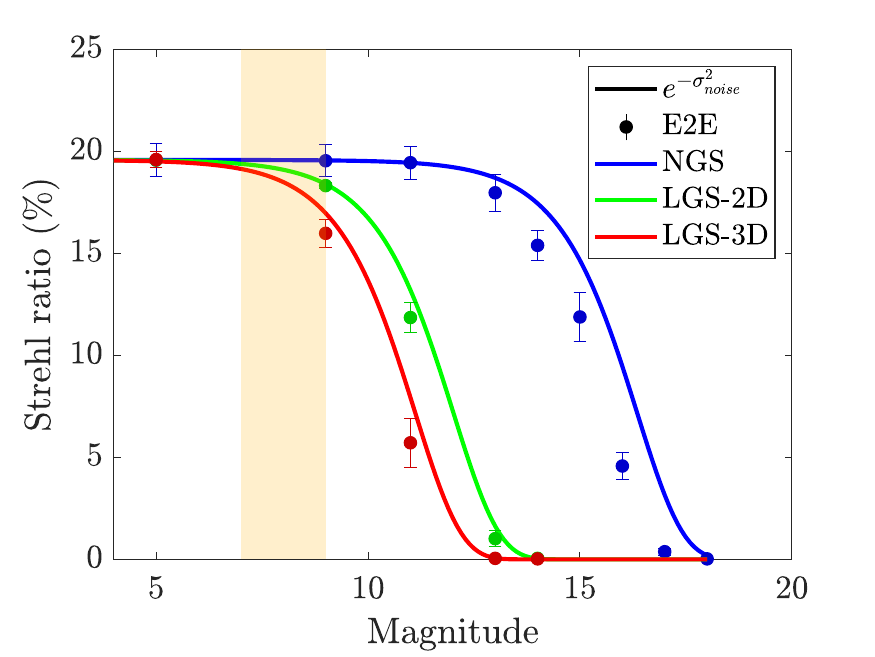}
    \caption{Strehl ratio for E2E simulations (markers) and sensitivity predictions (solid lines) of closed-loop performance for 8 (top plot) and 16 m (bottom plot) telescopes. $\sigma^{2}_{noise}$ corresponds to the residual variance predicted by fitting error and photon noise for NGS, LGS-2D and LGS-3D. The yellow stripe corresponds to the expected return flux of the laser guide stars for the ELT.}
    \label{fig:cl816}
\end{figure}

Due to computational limitations, we were able to perform E2E simulations of the closed loop for a maximum telescope diameter of 16 m. For this reason, we tested the predictive capabilities of the sensitivity method by simulating the E2E closed loop for 8 and 16 m telescopes. These results are shown in Fig. \ref{fig:cl816}, where the solid lines correspond to the theoretical performance predicted using the sensitivity analysis, including the fitting error, and the markers correspond to 20 realizations for different atmospheric phase screens. The top plot corresponds to the simulation using the 8 m telescope and the bottom plot to the 16 m. A vertical yellow stripe was added from magnitudes 7 to 9, to represent the expected return fluxes for the laser guide stars for the ELT.

The first result from these figures is that it is possible to close the loop for the extended objects using the calibration procedure utilizing the convolutional model and applying optical gains to optimize the point source calibrated interaction matrix and be able to obtain almost the same performance at high flux as the NGS. This indicates that by introducing a gain scheduling camera it would be possible to compute these optical gains and to have an online method of optimizing the reconstructor that follows the changing density profile of the sodium layer.

Given its higher sensitivity, an NGS can be used for up to three magnitudes more than for the extended objects for the 8 m telescope and for four to five magnitudes more in the 16 m telescope for the LGS-2D and LGS-3D respectively. In fact, the limiting magnitude for the NGS will only have a small to no dependence on the diameter of the telescope used. This is because, even though bigger telescopes collect more light, they also need more actuators to maintain the same actuator pitch, and these two effects cancel each other. For the LGS-2D, the increase in light collected gets almost exactly canceled by the increase in the relative size of the laser width which decreases the sensitivity, but as bigger telescopes need more actuators, increasing the diameter of the telescope decreases the limiting magnitude. For the LGS-3D the effect of the increase in relative size of the laser width and the increasing extension of the source makes the limiting magnitude decrease faster than for the LGS-2D. This effect can be seen in figure \ref{fig:cl816}, as the two curves for the extended object separate and move to the left.

When comparing the E2E results with the sensitivity analysis, it is possible to observe that the noise models accurately predict the performance of the closed loop system. What is interesting to note is that the computation time needed to obtain these solid curves is thousands of times less than the full E2E method to obtain the markers, as we can use the convolutional model to compute the sensitivity of the PWFS using extended objects.

\subsection{Extrapolating to 40 m}

\begin{figure}
    \centering
    \includegraphics[width = 0.49\textwidth]{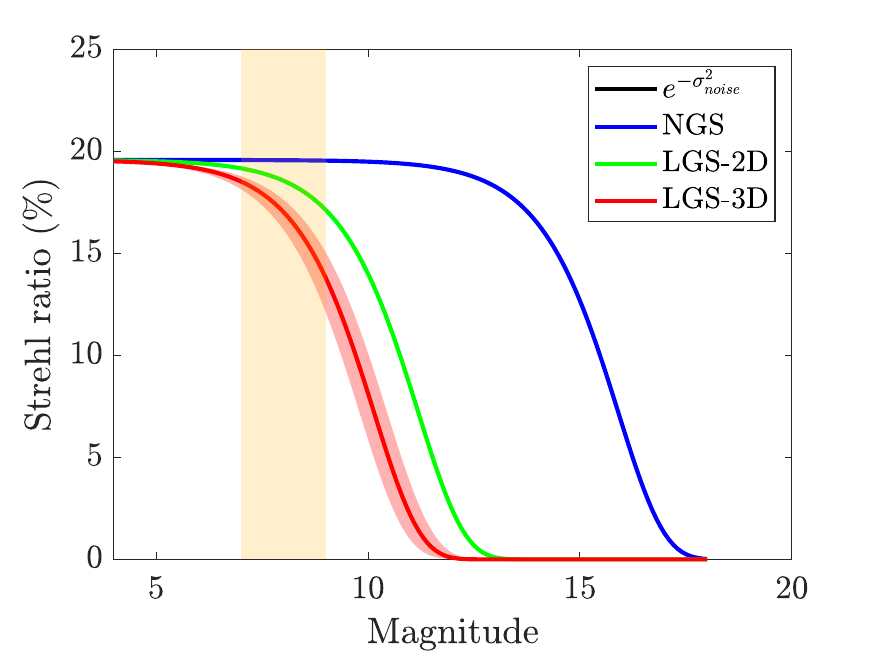}
    \caption{Sensitivity predictions of closed-loop performance for a 40 m telescope. For the LGS-3D the solid line corresponds to the average performance across all sodium profiles and the shaded region is limited by the maximum and minimum performance given all sodium profiles.}
    \label{fig:cl40}
\end{figure}

Now that we have shown that the sensitivity analysis can be used to predict the performance of the E2E closed-loop simulations, which would otherwise take hundreds or even thousands of hours to compute, we can now predict the expected performance for the 40 m telescope. To do this, we first built the interaction matrix for the 40 m telescope for an NGS for 5000 KL modes. This process, even though it was slow could be computed in a matter of hours. Then, we computed the reference intensity and a single focal plane image of the LGS as seen through the 40 m telescope. With the focal plane image, we could compute the optical gains using the convolutional model and then obtain the interaction matrix optimized for the LGS. With the calibration ready, we could get the sensitivities and compute how noise would propagate through the system. Fig. \ref{fig:cl40} shows the expected performance of the PWFS for the three tested sources for a 40 m telescope. In this plot it is possible to observe that the limiting magnitude of the NGS remains approximately constant with respect to the 8 and 16 m telescope cases, with a limiting magnitude 4.5 to 6.1 higher than for the extended objects. The difference in limiting magnitude for the LGS-2D and LGS-3D is 1.6 magnitudes and both extended objects have a drop in performance at around magnitude 8-9. As stated before, the simulated throughput of the system is 100 \%, but if taking 25 or 50 \%, then both extended objects will be operating in a condition where slight changes in the return flux of the LGS will mean large drops in performance, with the LGS-3D the most affected. 

It is interesting to note that even for the 8 m telescope (e.g. the VLT) the performance of the AO loop starts to drop at the expected return fluxes. This result implies that it might not be advisable for the VLT or ELT to use these kinds of wavefront sensors (knife-edge like WFS) with extended guide objects such as an LGS, given that the size of these objects reduces the sensitivity up to a point where the flux of the LGS is a limiting factor in the performance, as small variations in return flux might result in large drops in performance for both the LGS-2D and LGS-3D, the latter being the most affected. As we only tested for photon noise, this is a fundamental limit for the performance. If adding read-out noise, even if it has a small contribution, it will lower even further the performance.

\section{Conclusion}

In this work, we computed the expected performance of the AO loop for a PWFS using an LGS for a 40 m telescope. To do this, we used a sensitivity model to predict the residual phase due to photon noise, which when combined with control theory could predict the residual variance in closed-loop operation. For this model, we needed access to the interaction matrix calibrated for each source. 

To compute end-to-end interaction matrices, we introduced a new way to discretize an LGS and developed a new method on how to simulate extended objects for any FFWFS, which we called the ROI Propagation. With this method we could simulate the light propagation and obtain the signal of the wavefront sensor in a fraction of the time compared with traditional methods, yielding the same results and maintaining the E2E nature of the full field of view propagation. As even with this new method we could not simulate the 40 m telescope, we used it to show that it was possible to compute the interaction matrix for a point source, and then use optical gains to optimize it for the extended object. This procedure is not only useful for simulation purposes but also for a real telescope, where the calibration source will be most likely a point source.

To obtain these optical gains we proposed the use of a gain scheduling camera, which by using a convolutional model could use focal plane images of the source to compute the optical gains needed to optimize the interaction matrix for the extended object. We showed that the convolutional model accurately predicted the value of these optical gains, by comparing the values obtained using E2E simulations and the ones obtained using the model.

Finally, we performed simulations of a closed loop for 8 and 16 m telescopes and determined that it was possible to close the loop using the optimized point source calibrated interaction matrix. Also, we found good agreement between the results obtained using the E2E methods and the sensitivity model. With this, we could predict the performance of the PWFS when using an extended guide star for a 40 m telescope. We found that for both the LGS-2D and the LGS-3D the loss in sensitivity means that they will be operating in a region where the flux of the LGS will generate a drop in performance. Small variations in the return flux of the LGS will result in large variations of performance, an effect that would also happen for smaller telescopes.

An interesting alternative would be to design a translation invariant WFS, such that the size of the source would have a minor impact on the performance. With a higher sensitivity, it would be possible to run the loop at higher frequencies or observe at smaller wavelengths. A translation-invariant WFS could be, for example, a repeating phase mask, in which a specific pattern is repeated in space. As the pattern repeats, the position of each sample (and therefore the size of the source) will have a minimum impact. An issue with it would be the diffraction effects and therefore the possible loss of light.

With this, we conclude that the use of knife edge-like wavefront sensors might not be a good alternative for LGS wavefront sensing for 8 to 40 m telescopes, as even if the instrument is capable of dealing with the Z-extension of the source, the width of the laser beam is enough to lower the sensitivity such that photon noise decreases the performance of the AO loop considerably at the expected return flux of the LGS.

%--------------------------------------------------------------------

\begin{acknowledgements}
This work benefited from the support of the the French National Research Agency (ANR) with \emph{WOLF (ANR-18-CE31-0018)}, \emph{APPLY (ANR-19-CE31-0011)} and \emph{LabEx FOCUS (ANR-11-LABX-0013)}; the Programme Investissement Avenir \emph{F-CELT (ANR-21-ESRE-0008)}, the \emph{Action Spécifique Haute Résolution Angulaire (ASHRA)} of CNRS/INSU co-funded by CNES, the \emph{ECOS-CONYCIT} France-Chile cooperation (\emph{C20E02}), the \emph{ORP-H2020} Framework Programme of the European Commission’s (Grant number \emph{101004719}), \emph{STIC AmSud (21-STIC-09)}, the french government under the \emph{France 2030 investment plan}, as part of the \emph{Initiative d'Excellence d'Aix-Marseille Université A*MIDEX, program number AMX-22-RE-AB-151}, the \emph{Conseil régional Provence-Alpes-Côte d'Azur} with the \emph{emplois jeune doctorant} program, co-funded by \emph{First Light Imaging}, and the Millennium Science Initiative Program (ACIP, NCN19 161).
\end{acknowledgements}

\bibliographystyle{aa}
\bibliography{aanda}

\newpage

\appendix

\section{Signal from the PWFS}
\label{app:framework}

Let $I(\phi)$ be the intensity recorded in the detector of the pyramid when a phase $\phi$ is introduced to the system. The reduced intensity is defined as 

\begin{equation}
    \label{eq:red_int}
    \Delta I(\phi) = \frac{I(\phi)}{N_{ph}} - I_0,
\end{equation}

where $N_{ph}$ corresponds to the number of photons in the frame, and

\begin{equation}
    \label{eq:RI}
    I_0 = \frac{I(\phi = 0)}{N_{ph}},
\end{equation}

known as the reference intensity, which here is chosen to be the PWFS signal for a flat wavefront. Then, having a modal basis $[\phi_1, \dotsc, \phi_n]$ (KL modes in this work) it is possible to build an interaction matrix $\mathcal{D} = [\delta I(\phi_1), \dotsc, \delta I(\phi_n)]$, where 

\begin{equation}
    \delta I(\phi_i) = \frac{I(\epsilon \phi_i) - I(-\epsilon \phi_i)}{2\epsilon}
    \label{eq:pushpull}
\end{equation}

corresponds to the push-pull operation with $\epsilon$ small enough to remain in the linear regime of the sensor.

\section{Noise propagation}
\label{app:noise}
The residual variance due to read-out noise and photon noise introduced each time the PWFS makes a measurement of the wavefront can be computed as

\begin{equation}
    \sigma^2_{\phi_i} = \frac{N_{sap} \, \sigma_{RON}^2}{N_{ph}^2 \, s^2(\phi_i)} + \frac{1}{N_{ph}\, s_\gamma^2(\phi_i)},
    \label{eq:sigma_phi}
\end{equation}

with $N_{sap}$ the number of subapertures, $\sigma_{RON}$ the standard deviation of the electronic noise in each pixel, $s$ the RON sensitivity and $s_\gamma$ the sensitivity to photon noise. These sensitivities are obtained using the columns of the interaction matrix from equation \ref{eq:pushpull} as

\begin{equation}
    s(\phi_i) = \sqrt{N_{sap}} \, \, \left| \left| \delta I(\phi_i) \right|\right|_2,
\end{equation}
and 
\begin{equation}
    s_\gamma(\phi_i) = \left| \left| \frac{\delta I(\phi_i)}{\sqrt{I_0}} \right|\right|_2,
    \label{eq:s_gamma}
\end{equation}

with $||.||_2$ the two norm. One thing to keep in mind is the fact that photon noise sensitivity is dependent on the illumination pattern of the pupils in the detector. Therefore, an approximation is made in the computation of the sensitivities, which assumes that we are working in the linear regime of the sensor, such that the illumination pattern is the one corresponding to a flat wavefront reaching the PWFS

\begin{equation}
    I(\phi) = I_0 + \Delta I(\phi) \approx I_0, \, \text{for  $\phi \ll 1$}.
\end{equation}

\section{Noise in closed loop}
\label{app:CTheory}
Consider a controller to be a discrete integrator in the feedback path with gain $\alpha$, a DM modeled as a Zero-Order Hold (ZOH) and a WFS as a ZOH with a time delay of one period $T$ with an additional time delay of one period for the computation of the signal. The negative of the loop transmission is

\begin{equation}
    -LT(s) = \left(\frac{1 - e^{-sT}}{sT}\right)^2 \, e^{-2sT} \, \frac{\alpha}{1 - e^{-sT}},
\end{equation}

with $s$ the Laplace's transform variable that can be expressed as $s = j\omega$ with $j$ the imaginary unit and $\omega$ the angular frequency, for the purposes of computing the gain of the system.

As the forward path for the noise is equal to the loop transmission, the Noise Transfer Function $NTF(s)$ can be expressed as

\begin{equation}
    NTF(s) = \frac{LT(s)}{1 - LT(s)}.
    \label{eq:NTF}
\end{equation}

As photon noise is white noise it has uniform Power Spectral Density (PSD), therefore we can integrate the magnitude squared of the NTF over the bandwidth to obtain the total noise that is propagated through the AO loop

\begin{equation}
    \sigma^2_{noise} = \frac{\sigma_{\gamma}^2}{F} \, \int_{-F/2}^{F/2} \left|NTF(s)\right|^2_{s = j2\pi f} \, df.
    \label{eq:NTF_int}
\end{equation}

\section{Geometry of the LGS}
\label{app:LGS_geom}

Approximating the sodium layer to be plane-parallel, starting at a height $h_l$ and ending at $h_h$ ($h_l < h_h$), pointing the telescope at a zenith angle $\theta$ it's possible to compute that the approximate angular size $\Delta \alpha$ of a side-launch LGS as

\begin{equation}
    \Delta \alpha = \frac{D}{2} \cos \theta \left(\frac{1}{h_l} - \frac{1}{h_h}\right).
\end{equation}

Then, considering an effective focal length $f$, the extension normal to the focal plane $\Delta z$ can be computed as

\begin{equation}
    \Delta z = \frac{h_l \sec \theta \, f}{h_l \sec \theta - f} - \frac{h_h \sec \theta \, f}{h_h \sec \theta - f}.
\end{equation}

Assuming that the effective focal length of the telescope is much smaller than the distance to the sodium layer, we can perform a Taylor approximation of the denominator

\begin{equation}
    \Delta z \approx \left(f + \frac{f^2}{h_l}cos\theta\right) - \left(f + \frac{f^2}{h_h}cos\theta\right),
\end{equation}

therefore, the approximate expression for the normal extension of the LGS is

\begin{equation}
    \Delta z = f^2\,\cos\theta \left(\frac{1}{h_l} - \frac{1}{h_h}\right).
\end{equation}

\end{document}